\makeatletter
\g@addto@macro\normalsize{%
  \setlength\abovedisplayskip{3pt}
  \setlength\belowdisplayskip{3pt}
  \setlength\abovedisplayshortskip{2pt}
  \setlength\belowdisplayshortskip{2pt}
} \makeatother

\documentclass[useAMS,referee,usenatbib]{biom}
\usepackage{graphicx}
\usepackage{float,subfigure}
\usepackage{mathrsfs,amsmath}
\usepackage{amssymb} 
\usepackage{gensymb} 
\usepackage{multirow}
\usepackage{etoolbox}

\BeforeBeginEnvironment{figure}{\vskip-2ex}
\AfterEndEnvironment{figure}{\vskip-2ex}

\graphicspath{{./figs/}} 

\def \msb{\!_}
\newcommand{\nc}{\newcommand}
\nc{\nn}{\nonumber} \nc{\fns}{\footnotesize}

\nc{\revisionline}{\vspace{.1in} \today \vspace{.1in}
\hrule\hrule\hrule\vspace{.1in}} \nc{\newpp}{\vspace{.1in}
\noindent}

\nc{\slideline}{\smallskip \hrule\hrule \smallskip}

\nc{\wh}{\widehat} \nc{\wt}{\widetilde}

\nc{\bgam}{\pmb{\gamma}} \nc{\bgamma}{\pmb{\gamma}}
\nc{\bGamma}{\pmb{\Gamma}}

 \nc{\beq}{\begin{eqnarray*}}
\nc{\eeq}{\end{eqnarray*}}

\nc{\beqna}{\begin{eqnarray}} \nc{\eeqna}{\end{eqnarray}}

\nc{\lsq}{\left[} \nc{\rsq}{\right]} \nc{\lbc}{\left \{ }
\nc{\rbc}{\right \} } \nc{\lp}{\left(} \nc{\rp}{\right)}

\nc{\imp}{\Rightarrow} \nc{\lbf}{\lim_{b \rightarrow \infty}}
\nc{\limNinf}{\lim_{N \rightarrow \infty}} \nc{\limminf}{\lim_{m
\rightarrow \infty}} \nc{\limninf}{\lim_{n \rightarrow \infty}}
\nc{\convd}{\stackrel{\text{D}}{\longrightarrow}}
\nc{\convp}{\stackrel{\text{P}}{\longrightarrow}}
\nc{\convqm}{\stackrel{\text{qm}}{\longrightarrow}}
\nc{\eqd}{\stackrel{{\EuScript D}}{=}}
\nc{\convas}{\stackrel{a.s.}{\longrightarrow}}
\nc{\subi}{_{\text{I}}} \nc{\subs}{_{\text{S}}}
\nc{\subni}{_{\text{NI}}}
\nc{\trans}{^\top} \nc{\ol}{\overline}
\nc{\Ef}{ {\rm E}_{\infty} } \nc{\Ex}{ {\rm E} } \nc{\Ec}{ {\rm E}_1
} \nc{\Pf}{ {\rm P}_{\infty} } \nc{\Pc}{ {\rm P}_{1} } \nc{\Prb}{
{\rm P} } \nc{\sd}{\pm \hat{\sigma} }

\nc{\cond}{{\, \vert \,}} \nc{\indep}{{\, \perp \! \! \! \perp  \,}
} \nc{\tsps}{^{ {\rm T} } }

\nc{\pu}{\pi_{\rm U}} \nc{\pbi}{\pi_{\rm B}} \nc{\pnb}{\pi_{\rm NB}}
\nc{\prp}{\propto} \nc{\pr}{ {\rm pr} }
\nc{\half}{ {\textstyle \frac{1}{2}} }

\title[]{Modeling the Causal Effect of Treatment Initiation Time on Survival: Application to HIV/TB Co-infection}

\author{Liangyuan Hu$^{1,*}$\email{liangyuan.hu@mssm.edu}, 
Joseph W. \ Hogan$^{2}$, and Ann W.\ Mwangi$^{3}$, and Abraham Siika$^{3}$  \\
$^{1}$Icahn School of Medicine at Mount Sinai, New York, New York 10029, USA\\
$^{2}$Brown University School of Public Health, Providence, Rhode Island 02912, USA\\
$^{3}$Moi University School of Medicine, Eldoret 30100, Kenya
}

\begin{document}





\pagerange{\pageref{firstpage}--\pageref{lastpage}} 
\volume{64}


\doi{10.1111/j.1541-0420.2005.00454.x}


\label{firstpage}


\begin{abstract}
The timing of antiretroviral therapy (ART) initiation for HIV and tuberculosis (TB) co-infected patients needs to be considered carefully.  CD4 cell count can be used to guide decision making about when to initiate ART. Evidence from recent randomized trials and observational studies generally supports early initiation but does not provide information about effects of initiation time on a continuous scale.
In this paper, we develop and apply a highly flexible structural proportional hazards model for characterizing the effect of treatment initiation time on a survival distribution. The model can be fitted using a weighted
partial likelihood score function. Construction of both the score function and the weights must accommodate
censoring of the treatment initiation time, the outcome, or both.
The methods are applied to data on 4903 individuals with HIV/TB co-infection, derived from electronic health
records in a large HIV care program in Kenya. We use a model formulation that flexibly captures the joint
effects of ART initiation time and ART duration using natural cubic splines. The model is used to generate survival curves corresponding to specific treatment initiation times; and to identify optimal times for ART initiation for subgroups defined by CD4 count at time of TB diagnosis. Our findings potentially provide `higher resolution' information about the relationship between ART timing and mortality, and about the differential effect of ART timing within CD4 subgroups.
\end{abstract}

%



\maketitle


%

\section{Introduction}
\label{s:intro}

\subsection{Overview and Objectives}

Co-infection with HIV and tuberculosis (TB) is a significant public health
issue in the developing world.
Investigation of the most effective strategies
for integration of  antiretroviral therapy (ART)
for HIV with standard treatment for tuberculosis (TB) has 
been a topic of intensive research, including at least three
major randomized controlled trials (RCT) \citep{havlir2011timing, abdool2011integration, blanc2011earlier}.
Current World Health Organization (WHO)
guidelines recommend immediate initiation of TB therapy,
with concomitant or closely sequenced 
ART for HIV, depending on
presentation at diagnosis \citep{WHO2013}.

Initiating ART too
early increases the potential for drug toxicity and the risk of
TB-associated immune reconstitution inflammatory syndrome (IRIS),
a potentially fatal condition
\citep{havlir2011timing}.
On the other hand,  late initiation of ART increases risk for
morbidity and mortality associated with AIDS
\citep{abdool2010timing}.
Although optimal timing of ART initiation during the TB
treatment period is difficult to determine
\citep{abdool2011integration}, 
evidence from randomized trials strongly supports
earlier initiation for those who have advanced
HIV disease at the time of TB diagnosis, with more
equivocal conclusions for those at less advanced 
stages \citep{havlir2011timing, abdool2011integration, blanc2011earlier}.

Although  randomized trials form a strong evidence base,
limitations remain.  First, the randomized trials compare protocols defined
in terms of time {\em intervals} for ART initiation, rather than actual timing.
Second, the sample sizes are relatively small, with participants
distributed over multiple sites, in some cases in different countries
within the same trial.
Third, randomized trials are by necessity conducted under tightly controlled conditions,
and may not fully reflect the experience of patients in 
routine care settings.   
Data derived from electronic health records (EHR) provide an opportunity
to augment the evidence base using larger-scale data collected
in a representative care setting.  In particular, EHR enables 
investigation of the effect of ART timing at higher resolution because
actual initiation times are available and subgroups of interest have larger
sample sizes. However, any analysis of EHR and other types of observational data 
must address issues related to nonrandom treatment allocation, irregular sampling,
and complex missing data patterns.

\begin{figure}
 \centerline{%
 \includegraphics[scale=1]{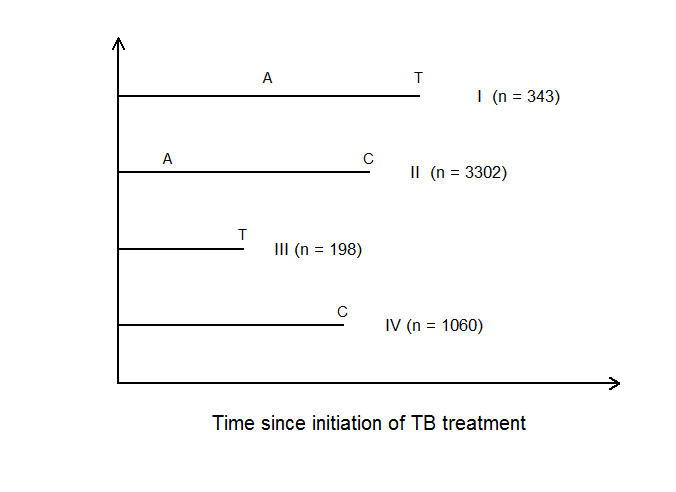}} 
\caption{Patterns of observed information on ART initiation time
$A$ and death time $T$. Either or both may be right censored prior
to one year at time $C$.}
\label{cenp}
\end{figure}

In this paper we develop a modeling framework
for drawing causal inferences about the timing of treatment
initiation when the outcome is an event time; in this case,
time to death.  
We apply the model to data drawn from the 
medical records of individuals enrolled in the Academic Model Providing Access
to Healthcare (AMPATH), a large-scale HIV care program in western
Kenya \citep{einterz2007responding}. 
The data have rich information, but the
observational nature of the data poses several complications for our analysis.
First, treatment initiation time
is not randomly allocated. 
Second, many individuals 
have incomplete information on 
exposure, outcome, or both.  
For example, some patients died before initiating ART,
which censors ART initiation time;
for others, incomplete follow up leads to censoring
of death time, ART initiation time, or both.
Figure~\ref{cenp} shows the four observed-data patterns
of ART initiation and mortality following initiation of  TB
treatment.  Third, the functional form of the causal effect
of initiation time on mortality rate is not known.\vspace{-1.2ex}
\vspace{-1.2ex}

To address these issues, we formulate a
structural causal hazard model that captures the effects of
both timing and duration of ART on hazard of mortality using unspecified smooth
functions, and we derive methods
for generating consistent estimates of model parameters
under nonrandom treatment allocation and  complex censoring patterns.
Using output from the fitted model, we generate 
estimates of the functional causal relationship
between  ART timing
and mortality.  Our method can be used to estimate the survival
distribution --- and therefore any functional of the survival distribution ---
associated with any specific ART initiation time.
Our analyses of the AMPATH data uses one-year survival
as the primary endpoint 
because it is consistent with one of the major randomized trials \citep{havlir2011timing}, but any functional of the survival
distribution could be used.

This paper is organized as follows: 
the remainder of Section 1 provides additional background
on issues related to HIV/TB
co-infection and reviews statistical approaches related
to estimating causal effect of treatment timing.
Section 2 describes
notation, defines optimal initiation time,
and provides detailed specification of our marginal
structural proportional hazards model. 
Section 3 develops the methods for
fitting the model to observational data where
both treatment timing  and death time are subject to censoring. 
Section 4 presents results from applying our analysis of AMPATH data,
with emphasis on new insights that can be gained,
relative to findings from randomized trials. Section 5 summarizes a simulation study that examines properties of model-based inferences.  In Section
6 we put the results in context by comparing them with RCT findings
and current treatment guidelines, and outline directions for further
research.

\subsection{Treatment of Individuals Co-infected with HIV and Tuberculosis}

Generally, treatment of TB itself is a 6-month regimen. 
The World Health Organization (WHO) recommends that TB
treatment should be initiated first,
with ART initiated
within the first 8 weeks for all patients
and within the first 2 weeks for those with CD4 less than 50.
However, complications
associated with early initiation of ART may discourage treatment
adherence and cause adverse effects
resulting from drug interactions and 
IRIS \citep{WHO2013}.  

Recent RCTs provide an evidence base
for these recommendations. For brevity, ART initiation time is hereafter defined relative to the start of TB treatment, and measured in weeks. 
The AACTG study A5221 found reduced 48-week mortality in the CD4$<$50 subgroup, comparing the early ($0-2$) to the late ($8-12$) initiation arm \citep{havlir2011timing}. 
The SAPIT trial recorded significantly higher mortality rate for the initiating interval $24-28$ compared to $0-4$ or $8-12$ \citep{abdool2011integration}.
In the CAMELIA trial, significantly reduced hazard of death was found in the early initiation arm (2 versus 8) \citep{blanc2011earlier}.

The question of ART timing has also been examined in 
at least {three observational studies.}
An analysis of  322 co-infected patients from Spain showed 
longer survival for those starting ART in the first 2
months of TB treatment, compared to 3 months \citep{velasco2009effect}. 
A prospective cohort study of 667 individuals in 
Thailand demonstrated an increased risk of death with 
longer delays in ART initiation \citep{varma2009hiv}. 
An analysis of 308 co-infected adults in Rwanda 
concluded that early ART initiation reduced mortality in
persons with CD4 count below 100
\citep{franke2011effectiveness}.

\subsection{Data Source:  AMPATH Medical Record System (AMRS)}  

AMPATH provides care to over 160,000 HIV-positive individuals at 143  sites
 in western Kenya.   The AMPATH Medical Record System
(AMRS), one of the largest of its kind in sub-Saharan Africa,
contains more than 100 million clinical observations from over 300,000 enrolled patients \citep{rachlis2015evaluating}. 
Our analysis makes use 
of clinical encounter data from 
4903 adults with HIV/TB co-infection drawn from the
AMRS between March 1, 2004 and April 18, 2008. The dataset contains individual-level information at baseline on the following variables: age, gender, WHO stage (a 4-level ordinal diagnostic indicator of HIV severity), weight, CD4 count, clinic visit (urban versus rural), marital status and post-primary education. The dataset includes also longitudinal information on ART initiation status, death, and CD4 count. 
Among  all patients presenting with HIV/TB co-infection, 26\% have incomplete information on ART initiation time (case III and IV in Figure~\ref{cenp}).

These data were generated before the findings from
the studies referenced above were released; this, combined
with the exercise of clinical judgment in making treatment
decisions, yields significant variation in ART initiation time, even
among those whose clinical profiles are similar in terms of
recorded information.  
\subsection{Statistical Methods for Treatment Duration and Timing}

Our methods address inference about causal effect
of treatment {\em initiation time} on a survival distribution.
The related problem of causal effect of treatment
{\em duration} (assuming initiation time is the same for everyone) has been investigated
in two papers by \citet{johnson2004estimating, johnson2005semiparametric}; 
the first considers the case where duration is discrete and the second
extends to the more complex case of continuous time.  In both cases,
duration time is subject to censoring that may be related to
measured covariates, and inverse probability weighting is used
to address nonrandom treatment allocation.  
Both papers restrict attention to the mean
or regression function related to a single endpoint; in our setting,
we use survival time as the endpoint, which introduces the need to deal 
with censoring of both the exposure and the endpoint.   
Our approach to inference builds on 
\citet{johnson2004estimating, johnson2005semiparametric}, 
particularly as it relates to the use of
a Radon-Nikodym derivative for deriving appropriate weight functions
(see also \citealp{murphy2001marginal}). Work by \cite{xiao2014flexible} uses regression splines to model the weighted sum of treatment duration and estimate the effect of cumulative exposure to ART. Our model includes a term for treatment duration as a function of time, separately captures effect of timing, and allows the effect of duration to depend on timing (provided sufficient data are available). 

 
The analysis by \citet{franke2011effectiveness}, referenced above,
uses a model that has a similar formulation to ours.  They use
pooled logistic regression with inverse probability weights
to fit a marginal structural model to a sample from 308 individuals.
Our paper formalizes that approach for continuous-time settings
and considerably relaxes distributional assumptions about the underlying
potential outcomes distributions.  

The fitted model can be used
to estimate the entire distribution of potential outcomes 
corresponding to a given initiation time.  The estimated causal
relationships are highly data-driven in the sense that 
key parts of the structural model are left unspecified, including the
baseline hazard function, the instantaneous effect of treatment
as a function of initiation time, and the effect of treatment
duration once initiated.   The added versatility of our model
is motivated by the desire to take full advantage
of the information available in our much larger cohort
(which, to our knowledge, is larger than any cohort
associated with a published analysis on HIV/TB co-infection).

\section{Formulation of Potential Outcomes Model}\label{formulation}

\subsection{Notation for Potential Outcomes and Observed Data}\label{notation_p1} 

Let $t$ denote time elapsed from the
initiation of TB therapy, and let $t_{\rmn{max}}$ denote maximum
follow-up time.
 Let $T_a > 0$ denote the potential outcome
corresponding to time of death under the scenario that ART is initiated at time $a$,
where $a \geq 0$ is continuous. We use $T_\infty$ to denote death
time corresponding to any $a>t_{\rmn{max}}$.

Information on potential outcomes is observed according to the cases
depicted in Figure~\ref{cenp}.  Let $A$ be the random variable
representing ART initiation time and let $T = T_A$ denote the
survival time corresponding to initiating ART at $A$. 
Let $C$ denote a
censoring time.
As shown in
Figure~\ref{cenp}, either or both of $A$ and $T_A$ may be right
censored, for example due to loss to follow up.  The observed data available for drawing inference about the distribution of potential outcomes are as follows:  the observed follow up time for the mortality outcome is $T^* = \min(T,C)$, with event indicator $\Delta^T = I(T < C)$.
Similarly, the observed follow up time for ART initiation time is
$A^* = \min(A, T^*) = \min(A,T,C)$, with event indicator $\Delta^A
= I(A < T^*)$. Notice that $A$ can be right censored by $T$, as
shown in Case II of Figure~\ref{cenp}. Because large ART initiation times are infrequently observed in our data, 
we administratively censor $A$ and $T^*$ at $t_{\rmn{max}}$ for cases where 
$A \geq t_{\rmn{max}}$ and $T^* > t_{\rmn{max}}$, respectively. 

Each individual has $p$ covariates, some of
which may be time varying.  We use $L(t)$, $t\geq 0$ to denote the
$p \times 1$ vector containing the 
most recently observed value of each covariate at time $t$, with
$\ol{L}(t) = \{ L(s) : 0 \leq s < t \}$ representing the observed
covariate history up to but not including $t$.  For each individual,
therefore, we observe a copy of $\lbc \ol{L}(A^*), T^*, \Delta^T,
A^*, \Delta^A \rbc$. The letter $A$ was originally used for AZT (zidovudine) and $L$ for lymphocyte cell count in earlier causal inference literature (Miguel Hernan, personal communication). 

\subsection{Marginal Structural Model for Mortality}
We assume $T_a$ follows a marginal structural proportional hazards
model of the form 
\begin{eqnarray} \label{coxmsm_p1}
\lambda_a(t) &=&
\lambda_\infty(t) \; r(t,a), 
\end{eqnarray} 
where $\lambda_a(t)$ is
the hazard function for $T_a$, $\lambda_\infty(t)$ is the reference
hazard for $T_\infty$, and $r(t,a)>0$ is the 
(time-dependent) hazard ratio function.
We parameterize $r(t,a)$ in terms of three functions,
$g_1(a)$, $g_2(t-a)$ and $g_3(a(t-a))$, respectively denoting effects of timing, duration, and timing-duration interaction. 
Specifically, we assume 
\beqna \label{eq:rmodel}r(t,a) &=&  \exp  \lsq I(a < t) \lbc g_1(a) + g_2(t-a) + g_3(a (t-a))  \rbc \rsq,
\eeqna 
where $g_1(\cdot)$, $g_2(\cdot)$ and $g_3(\cdot)$ are unspecified smooth functions that are twice
continuously differentiable, with the second derivative set to zero
at the interval boundaries for $a$, $t-a$ and $a(t-a)$. 
The hazard function $\lambda_a(t)$ 
can depend on baseline covariates $L(0)$ by
elaborating model~(\ref{eq:rmodel}) or by using a stratified
version of $\lambda_\infty(t)$, as demonstrated in 
Section~\ref{sec:application}. 

\begin{figure}[H]
\centering
\subfigure[$g_1(a_1) = g_1(a_2) = -.5; \; g_2(a) =0$]{%
\includegraphics[scale=0.5]{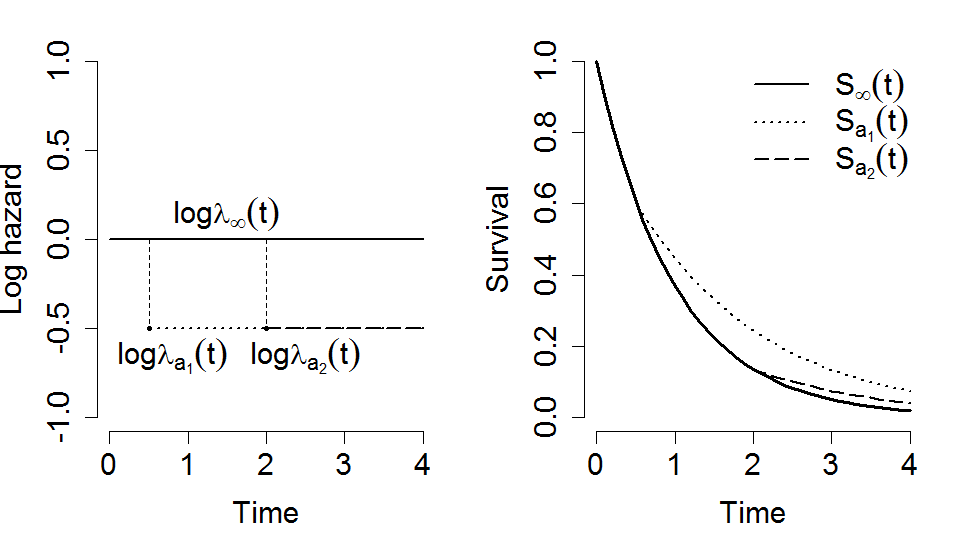}
\label{example 1}}
\subfigure[$g_1(a_1) = -1.5;\; g_1(a_2) = -.5; \; g_2(a) =0$]{%
\includegraphics[scale=0.5]{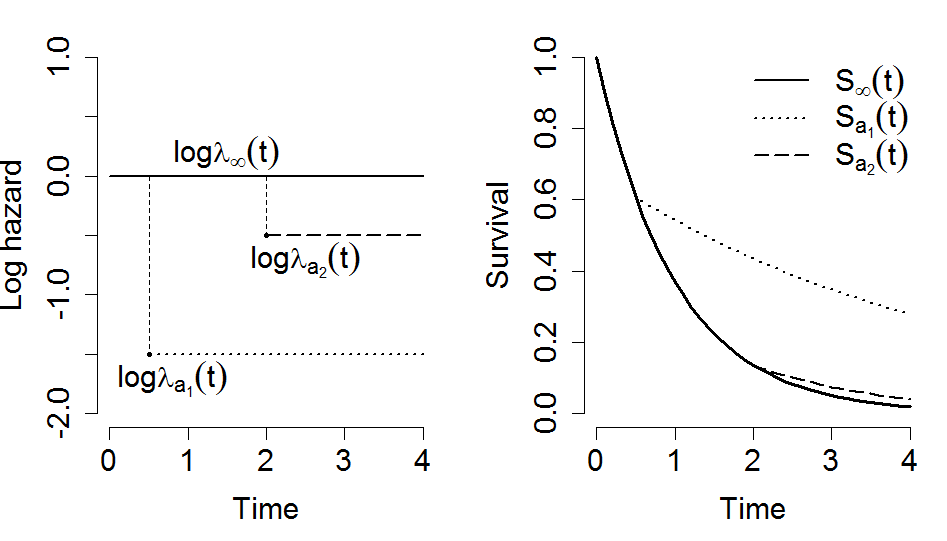}
\label{example 2}}
\subfigure[$g_1(a_1) = -1.5;\; g_1(a_2) = -.5; \; g_2(a) = -.5(t-a)$]{%
\includegraphics[scale=0.5]{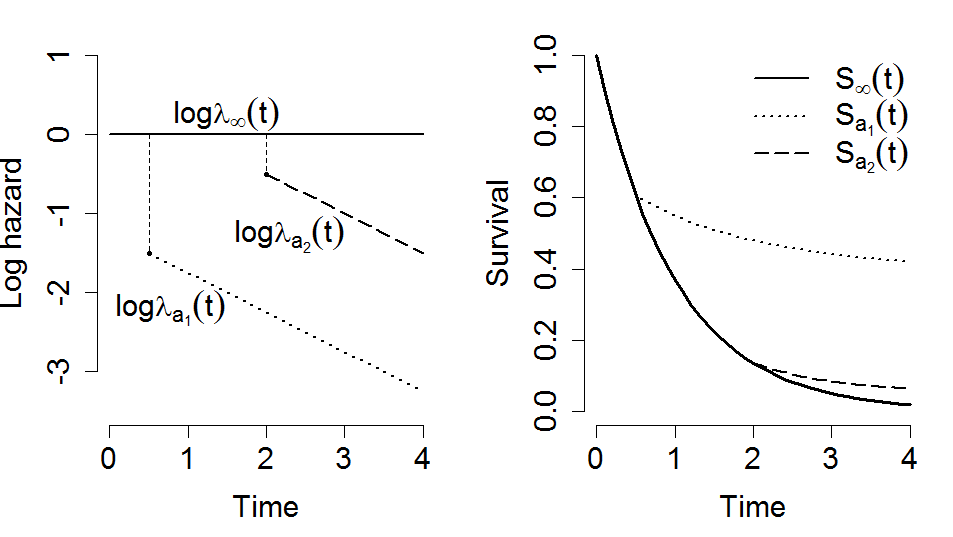}
\label{example 3}} 
\caption{Simple examples of the hazard models and
corresponding survival models; Two initialization times are
compared: $a_1 = .5; \; a_2 = 2.$} \label{hazmod.eg}
\end{figure}

The model can be understood by considering some simple cases. Figure~\ref{hazmod.eg}
illustrates three simplified versions of the hazard model and the
corresponding survival models. For clarity, models in
Figure~\ref{hazmod.eg} assume $\lambda_\infty(t)$ is constant
(our model allows it to remain unspecified)
and assumes $g_3(a(t-a))=0$ (no interaction between $a$ and $t-a$). 
In Figure 2(a), we
set $g_1(a) = \beta_1$ and $g_2(t-a)=0$,
which implies
the log hazard of mortality is $\log \{ \lambda_\infty(t) \}$ prior
to treatment initiation ($t<a$) and $\beta_1 +  \log \{
\lambda_\infty(t) \}$ thereafter ($t \geq a$). This version
assumes the instantaneous effect of initiating ART on hazard of
mortality is not a function of initiation timing. 
In Figure 2(b), we allow $g_1(a)$ to vary with $a$, so that the effect
of ART initiation does depend on its timing.  In Figure 2(c),
we additionally set $g_2(t-a) = (t-a) \beta_2$, where $\beta_2$
is a scalar, which implies that the instantaneous effect of
initiating ART is $g_1(a)$, and that the effect of {\em being on}
ART at any given time $t$ depends on duration. 

 \vspace{-1.2ex}
\vspace{-1.2ex}

Turning to the more general formulation, we parameterize  $g_1(\cdot)$, $g_2(\cdot)$ and $g_3(\cdot)$  using natural cubic
splines constructed from piecewise third-order polynomials that pass
through a set of control points, or knots. 
A natural cubic spline has continuous first and second derivatives at
the knots, and is linear beyond the boundary knots. Basis functions
generated under these constraints are called B-spline functions
\citep{hastie2009elements}. Parameterizing~\eqref{eq:rmodel} in terms of
B-splines yields 
\beqna \label{eq:hazardmodel3_p1} 
\lambda_a(t) &=&\lambda_{\infty}(t)\exp \lsq I(a<t)\lbc 
 b_1\trans(a) \beta_{1} + b_2\trans(t-a) \beta_{2} +  b_3\trans(a  (t-a)) \beta_{3} \rbc\rsq,
\eeqna where $b_1(\cdot) =
(b_{\msb{11}}(\cdot),\ldots,b_{\!_{1K_1}}(\cdot))\trans$ is a
$K_1 \times 1$ vector representing a 
B-spline basis function of degree $K_1$ in $a$, and 
$b_2(\cdot) =
(b_{\msb{21}}(\cdot),\ldots,b_{\msb{2K_2}}(\cdot))\trans$ and $b_3(\cdot) =
(b_{\msb{31}}(\cdot),\ldots,b_{\msb{3K_3}}(\cdot))\trans$,
having dimension $K_2 \times 1$ and $K_3 \times 1$, respectively, are 
defined similarly as functions of $t-a$ and $a(t-a)$. The parameters
$\beta_{1}$, $\beta_{2}$ and $\beta_3$ are, respectively,
$K_1 \times 1$, $K_2 \times 1$ and $K_3 \times 1$ vectors 
 of coefficients for  the basis functions $b_1(\cdot)$, $b_2(\cdot)$ and $b_3(\cdot)$.
The model for the risk function can be written
in compact form as  $r(t,a) = \exp \{X\trans (a,t)\beta\},$
where 
\beqna \label{Xdmat_p1} 
X(a,t)~ \msb{(K_1 + K_2  + K_3) \times 1} =
I(a<t)[ b_1\trans(a), b_2\trans(t-a), b_3\trans(a (t-a))]\trans \eeqna and 
$\beta ~\msb{(K_1 + K_2 +K_3) \times 1} = (\beta_1 \trans , \beta_2
\trans, \beta_3\trans)\trans$.  The causal effect of ART initiation time on
the survival distribution is therefore encoded in $\beta$,
and the formulation in terms of B-spline basis functions
allows the use of weighted proportional hazards regression
methods to obtain consistent estimates of $\beta$.

\section{Estimation of Structural Model and Optimal Initiation Time}\label{sec:estimation}

\subsection{Overview}

In our data, the decision to initiate ART depends on
information available to the physician, and is therefore not
randomly allocated. Moreover, $A$ will be right censored when an
individual dies or leaves the study before treatment initiation.  
We propose to estimate $\beta$ by maximizing an inverse-probability-weighted
partial likelihood score function.
We show how to construct and estimate weights that lead to 
consistent estimates of $\beta$ under specific assumptions about
treatment allocation and censoring, and in Web Appendix Section 1, we 
provide a heuristic justification  based on the use of 
Radon-Nikodym derivatives \citep{murphy2001marginal}.

In deriving the weighted estimating equations for $\beta$, 
we first consider the hypothetical case
where $A$ is not randomized and where $T$ is always observed.  Next,
we move to the case where $A$ may be censored by $T$ (i.e., where
death occurs prior to treatment initiation), but there is no
censoring by $C$. We then describe the case where $A$ is not
randomized, and possibly censored by $T$. Finally, we generalize our
approach to allow for censoring by $C$. 
We then  show how to construct causal contrasts
between different treatment initiation times using
output from the fitted model.

\subsection{Randomized Treatment Assignment}

If treatment is randomly allocated according to a known probability density $f^A(\cdot)$, 
we can use the standard partial likelihood score equations to derive
consistent estimators of $\beta$. 
For this simple case we assume $A$ is observed for everyone and
there is no censoring by $C$.
For $t>0$, let $N^T(t) = I\lbc T \leq t, \Delta^T = 1 \rbc$
denote the
zero-one counting process associated with $T$.
Let $Y(t)=1$ if an individual is
still at risk for death and under observation at $t$, and $Y(t) = 0$
otherwise. The partial likelihood score equations can be written
$\sum_{i=1}^n D_i(\beta) = 0$, where 
\beqna \label{eq:eeRCT}
D_i(\beta) &=& D(A_i,T_i; \beta) \nonumber\\
                     &=& \int_0^\infty \lbc X(A_i,t) - \ol{X}(t,\beta) \rbc dN^T_i(t),
\eeqna 
where $X(A_i,t)$ is the design matrix based
on~(\ref{Xdmat_p1}) and \beq \ol{X}(t,\beta) &=& \frac{\sum_k
X(A_k,t) Y_k(t) r(A_k,t; \beta) } { \sum_k Y_k(t) r(A_k,t; \beta)} .
\eeq Let $E_R ( \cdot )$ denote expectation under randomized
treatment assignment. Under randomization of $A$, $n^{-1} \sum_i
D(A_i,T_i; \beta)$ is an unbiased estimator of $E_R \lbc D(A,T;
\beta) \rbc$. The estimating function $\sum_i D_i(\beta)$ is a stochastic
integral of a predictable process with respect to a martingale, and
as such $E_R \lbc D(A,T; \beta_0) \rbc = 0$ at the true value
$\beta_0$ of $\beta$; hence, the root $\wh{\beta}$ of the 
estimating equation $\sum_{i=1}^n D_i(\beta) = 0$ is a consistent
estimator of $\beta$ \citep[pp. 297-298]{fleming2005counting}.

Now consider the case where $A$ is still randomly allocated, but
death may occur before initiation of treatment, so that $A^* =
\min(A,T)$ and $\Delta^A = I(A<T)$.  We continue to assume no
censoring by $C$. Following \citet{johnson2005semiparametric}, the
mean of an individual score contribution can be represented as the sum of contributions
from those with observed and censored values of $A$; i.e., 
$E_R \lbc
D(A_i,T_i;\beta) \rbc = E_R \lbc \Delta^A_i D(A_i,T_i; \beta) +
(1-\Delta^A_i) D(A_i,T_i; \beta) \rbc$. 
Contributions corresponding to censored values of $A$ have expectation
\beqna E_R \lbc
(1-\Delta^A_i) D(A_i,T_i; \beta) \rbc
&=& E_R \lsq E_R \lbc (1-\Delta^A_i) D(A_i,T_i; \beta) \, \left| \, \Delta^A_i, A^*_i \right. \rbc \rsq \nn \\
&=& E_R \lsq (1-\Delta^A_i) E_R \lbc D(A_i,T_i; \beta) \, \left| \, \Delta^A_i, A^*_i \right. \rbc \rsq \nn \\
&=& E_R \lbc(1 - \Delta^A_i) \int_{A^*_i}^\infty D(a,T_i,\beta) dF^{A \cond A>A^*_i}(a) \rbc \nn \\
&=& E_R \lbc \frac{(1 - \Delta^A_i)}{1 - F^A(A^*_i)}
                                \int_{A^*_i}^\infty D(a,T_i,\beta) dF^{A}(a) \rbc. \label{eq:int}
\eeqna 
To evaluate~(\ref{eq:int}), note that $D(a,T_i; \beta) = \int_0^\infty \{ X(a,t) -
\ol{X}(t,\beta) \} dN^T_i(t)$, and recall that $X(a,t) = I(a<t)[ b_1\trans(a),
b_2\trans(t-a), b_3\trans(a(t-a))]\trans$, 
so that $X(a,t) = 0$ for $a \geq t$.
This implies $D(a,T_i,\beta)$ is a constant function of $a$,
with  $D(a,T_i,\beta)= -\ol{X}(T_i,\beta)$ 
for $a \geq A_i^*$ and 0 otherwise.
The integral in~(\ref{eq:int}) therefore reduces to
$-\ol{X}(T_i,\beta) \{ 1-F^A(A^*) \}$, and the expectation itself
reduces to $E_R \lbc - (1-\Delta^A_i) \ol{X}(T_i,\beta) \rbc$.

Pulling this all together, we see that 
when death may occur before treatment initiation, and when there is
no censoring by~$C$, $n^{-1} \sum_i D_i(\beta)$ is an unbiased
estimator of\\ $E_R \lbc \Delta^A D(A,T; \beta) - (1-\Delta^A)
\ol{X}(T,\beta) \rbc$,
which implies that the solution $\wh{\beta}$
to the modified estimating equations \beqna \sum_{i=1}^n \lbc
\Delta_i^A D(A_i,T_i; \beta) - (1-\Delta_i^A) \ol{X}(T_i,\beta) \rbc
&=& 0 \label{ee:nocens} \eeqna will yield a consistent estimator of
$\beta$.


\subsection{Non-random Allocation of Treatment}
Suppose now that $A$ is not randomly allocated, but that treatment
allocation can be considered ignorable in the sense that 
\beqna
\label{assum:sr_p1}    
\lambda^A ( t \cond \ol{L}(t), \mathscr{T}_{ \{a \geq t \} } )
			&=& \lambda^A (t \cond \ol{L}(t) ), 
\eeqna 
where $\mathscr{T}_{ \{ a \geq t \} } = \{
T_a : a \geq t \}$ is the set of potential failure times associated
with initiation times beyond $t$. This assumption states that
initiation of treatment at time $t$ is sequentially randomized in
the sense that it is independent of future potential outcomes,
conditionally on observed covariate history $\ol{L}(t)$
\citep{jambs1999marginal}. 

Let $\mathbb{P}_R(\cdot)$ denote the data distribution under
randomized treatment, and let $\mathbb{P}_O(\cdot)$ denote the same
under non-random allocation of treatment. Recall that the {\em observable}
data for each individual,   under
either randomized or non-randomized allocation of treatment, is
$\{A^*, \Delta^A, T^*, \Delta^T \}$. Following
\citet{murphy2001marginal} and Johnson and Tsiatis (2005), under the
sequential randomization assumption in~(\ref{assum:sr_p1}) and some
regularity conditions, including the positivity assumption referenced in Web Appendix Section 1,
the distribution of $\{A^*, \Delta^A, T^*, \Delta^T \}$ under
$\mathbb{P}_R(\cdot)$ is absolutely continuous with respect to the
distribution of $\{A^*, \Delta^A, T^*, \Delta^T \}$ under
$\mathbb{P}_O(\cdot)$, and a version of the Radon-Nikodym (R-N)
derivative is 
\beqna
E_O 
\lbc 
\left.
	\frac{\Delta^A \, f^A(A^*)}{f^A(A^* \cond \ol{L}(A^*)) } 
	+  \frac{(1-\Delta^A)(1- F^A(A^*))}{1 - F^A(A^* \cond \ol{L}(A^*))} 
\, \right| \, 
A^* = a, \Delta^A = \delta^A, T^* = t, \Delta^T = \delta^T
\rbc. \label{eq:RN}
\eeqna

An estimating equation
that is a function of observed data and is unbiased under the
distribution of $\mathbb{P}_R(\cdot)$ can be re-weighted by the R-N
derivative to obtain an unbiased estimating equation using the same
observed data, but now under the distribution $\mathbb{P}_O(\cdot)$ 
\citep{murphy2001marginal}. 
Define weights $W_{1i}^A$ and $W_{2i}^A$ as
\beq 
W_{1i}^A(t) &=& \frac{f^A(t)}{f^A(t \cond \ol{L}_i(t))
}, \; \; \; W_{2i}^A(t) \; = \; \frac{1 - F^A(t)}{1 - F^A(t \cond
\ol{L}_i(t)) }. \eeq  The R-N derivative in~(\ref{eq:RN}) suggests
using the weighted estimating equation \beqna \label{eq:wplnoc}
\sum_{i=1}^n \Delta_i^T\lbc \Delta_i^A D^*(A_i, T_i; \beta) W_{1i}^A(A_i)
                        - (1-\Delta_i^A) \ol{X}^*(T_i; \beta) W_{2i}^A(T_i)\rbc &=& 0,
\eeqna where $D^*$ and $\ol{X}^*$ are evaluated using weighted risk
set indicators \beq Y^*_i(t) &=& Y_i(t) \lbc I(A_i < t)
W_{1i}^A(A_i) + I(A_i \geq t) W_{2i}^A(t) \rbc. \eeq

Up to now we have assumed no censoring by $C$. 
Let  $N^A(t) = I\{ A \leq t, \Delta^A=1 \}$, for $t>0$,
denote
the zero-one counting process for treatment initiation, with
 $\ol{N}^A(t)$ representing treatment history information up to time $t$. 
Similarly, define $N^C(t) = I(C \leq t, \Delta^T=0)$.
We assume that censoring at $t$ can depend on covariate and treatment history,
but conditionally on these is independent of future potential outcomes.
This assumption can be expressed formally in terms of the hazard
function associated with $N^C(t)$, 
\beqna 
\label{assum:sr.c} 
\lambda^C ( t \cond \ol{N}^A(t), \; \ol{L}(t), \; \mathscr{T}_{ \{a
\geq t \}} ) 
&=& \lambda^C ( t \cond \ol{N}^A(t),\;
\ol{L}(t) ). 
\eeqna 
As with the treatment initiation process, we can define
a weight function associated with censoring,  \vspace{-8mm}
\beq 
W^C_i(t) &=& \frac{1- F^C (t)}{1- F^C(t \cond \ol{N}^A(t), \ol{L}_i (t))}. 
\eeq 
This leads to a final modification of the estimating equations
for $\beta$ to accommodate covariate- and treatment-dependent censoring,
\beqna 
\label{eq:wplc}
\hspace{-8mm} U_{n}(\beta) &=& \sum_{i=1}^n \Delta_i^T ~ W^C_i(T_i) \lbc \Delta_i^A
D^{**}(A_i, T_i; \beta) W_{1i}^A(A_i)
                        - (1-\Delta_i^A) \ol{X}^{**}(T_i; \beta) W_{2i}^A(T_i)\rbc,
\eeqna 
where $D^{**}$ and $\ol{X}^{**}$ are evaluated using weighted
risk set indicators 
\beq Y^{**}_i(t) &=& Y_i(t) \, W_i^C(t) \lbc I(A_i < t)
W_{1i}^A(A_i) + I(A_i \geq t) W_{2i}^A(t) \rbc. 
\eeq
Writing~(\ref{eq:wplc}) in terms of counting process notation used in~(\ref{eq:eeRCT}) makes the distinct contributions of the four observation patterns
in Figure~\ref{cenp} more transparent,
\beq
U_n(\beta) &=&\sum_{i=1}^n
\int_0^\infty 
W^C_i(t)
\lsq
\Delta_i^A \, W_{1i}^A(A_i) \lbc X(A_i,t) - \ol{X}^{**}(t; \beta) \rbc 
- 
(1-\Delta_i^A) \, W_{2i}^A(t) \, \ol{X}^{**}(t; \beta) 
\rsq
dN_i^T(t).
\eeq
In particular all individuals  censored by $C$, either before
or after initiating treatment (cases II and IV in Figure 1), 
contribute information about $\beta$
via contributions to the risk set up to the observed censoring time $C$.

\subsection{Estimation of the Weights}\label{weights}

The weights in~(\ref{eq:wplc}) depend on the marginal and
conditional density functions of $A$ and $C$,
which can be estimated from fitted models of
the marginal and conditional
intensity processes associated with $N^A(t)$ and $N^C(t)$.
To estimate $f^A(t \cond \ol{L}(t))$, we  assume
$\lambda^A(t \cond \ol{L}(t))$ follows a proportional
hazards regression parameterized using a finite-dimensional
parameter $\alpha^A$,
\beqna
\label{Ahazmod.d}
\lambda^A  ( t \cond \ol{L}(t)) &=& 
	\lambda_0^A(t) \, r^A(\ol{L}(t); \,\alpha^A),
\eeqna
where $r$ is a user-specified regression function.
(We use a proportional hazards formulation for convenience, but
any regression formulation can be used here). 
The conditional density function $f^A (t \cond \ol{L} (t) )$ can be estimated using 
the empirical cumulative hazard,
\beq 
\Lambda^A ( t \cond \ol{L}(t), \wh{\alpha}^A ) &=& \int_0^t \dfrac{r^A ( L(s); \, \wh{\alpha}^A ) d N(s)}{\sum_j Y_j(s) r^A  ( L(s); \, \wh{\alpha}^A )}  \\
 1- F^A(t \cond \ol{L}(t),\wh{\alpha}^A) &=& \exp \lbc -\Lambda^A ( t \cond \ol{L}(t), \wh{\alpha}^A )  \rbc, \\
 \wh{f}^A (t \cond \ol{L}(t) )  & = & \lambda^A(t \cond
\ol{L}(t), \wh{\alpha}^A ) \lbc 1- F^A(t \cond
\ol{L}(t),\wh{\alpha}^A)\rbc, \eeq   where $\wh{\alpha}^A$ is the
maximum partial likelihood estimator for model~(\ref{Ahazmod.d}).

To estimate the unknown marginal probability density $f^A(t)$, we
use the Nelson-Aalen estimator 
$\wh{\Lambda}^A(t) = \sum_i \int_0^t \, dN_i^A(s) / Y_i(s)$ for the
cumulative hazard function and  
$\wh{\lambda}^A(t) = d \wh{\Lambda}^A(t)$ for the hazard
function.  The estimated CDF and density functions are obtained 
using $\wh{F}^A(t)  =  1 - \exp \{ -\wh{\Lambda}^A (t) \}$
and $\wh{f}^A(t)  =  \wh{\lambda}^A(t) \{ 1- \wh{F}^A (t) \}$,
respectively.  The weight estimators are therefore a function
of $\wh{F}^A$, $\wh{f}^A$, and $\wh{\alpha}^A$, 
\beq \wh{W}_{1i}^A(t) &=& \frac{\wh{f}^A(t)}{\wh{f}^A(t \cond
\ol{L}_i(t))}, \; \; \; \wh{W}_{2i}^A(t) \; = \; \frac{1 -
\wh{F}^A(t)}{1 - F^A(t \cond \ol{L}_i(t), \wh{\alpha}^A) }. \eeq 
The weights $W_i^C (t)$ can be estimated in a similar fashion by 
specifying and fitting a regression model of the form $\lambda^C (t \cond \ol{N}^A(t), \, \ol{L}(t) )
= \lambda_0^C(t) \, r^C( \ol{N}^A(t), \, \ol{L}(t), \, \alpha^C)$
to obtain estimates of $f^C(t \cond \ol{N}^A(t), \ol{L}(t))$ and 
$F^C(t \cond \ol{N}^A(t), \ol{L}(t))$,
and using the Nelson-Aalen estimator \\
$\wh{\Lambda}^C(t) = \sum_i {\int_0^t} \, dN_i^C(s) / Y_i(s)$
to obtain estimates of $f^C(t)$ and $F^C(t)$.

\subsection{Optimal Initiation Time}\label{opttime}

Referring back to our structural hazard model~(\ref{eq:rmodel}), the 
survival function for the potential outcome $T_a$ corresponding
to initiation time $a$ is $S_a(t) =
\exp\{-\Lambda_a(t)\}$, where 
\beqna
\Lambda_a(t)
&=& I(t<a) \Lambda_\infty(t) \nonumber \\ 
&+& I(t\geq a) \lsq \Lambda_\infty (a) + \int_a^t \exp\{g_1(a) +
g_2(u-a) + g_3(a(u-a)) \}\mathrm{d}\Lambda_\infty(u) \rsq.
\label{eq:surv} 
\eeqna
The $g$ functions are estimated using weighted partial likelihood
score as described above, and the baseline cumulative hazard 
$\Lambda_\infty(t)$ is
estimated using the (weighted) Breslow estimator that arises from the fitting process.
The survivor function $S_a (t)$ can therefore be estimated
for  any combination of $a$ and $t$, enabling estimation of specific causal
contrasts such as mortality ratios $S_a(t) / S_{a'}(t)$ for~$a \neq a'$.

With sufficient amounts of data, 
we also can use model output to infer optimal timing for treatment
initiation.
An optimal initiation time is defined as the value of $a$ that
maximizes an objective function written in terms of a functional of
the distribution of potential outcomes. Specifically, let $F_a(t) =
P(T_a \leq t)$ denote the CDF associated with $T_a$, and let
$\theta_a = \theta(F_a)$ denote a scalar functional of $F_a$
(e.g., the mean $\int t \; dF_a(t)$ or median 
$ F_a^{-1}(\half)$).  For a given functional
$\theta_a$, the optimal initiation time is 
$a_{\text{opt}} = \arg \max_a \theta_a$.  

In our application we use one-year survival
$\theta_a = 1-F_a(t_0)$, with $t_0$ set to 52 weeks,
as the primary endpoint.  Our estimates of $a_{\text{opt}}$
were somewhat unstable because the one-year mortality curve as a function of initiation time appears to be monotone increasing from zero (i.e., nonconvex). Alternatively, we compare mortality rates for contextually motivated initiation time intervals (trials referenced in Section 1) to draw inferences about optimal timing.
We elaborate in Section~\ref{sec:application}.

\section{Application to AMPATH data}\label{sec:application}

Our analysis dataset contains 4903 HIV/TB co-infected patients who had TB therapy initiated and had a CD4 count below 350 at the start of TB therapy.  Under guidelines in place at the time, these patients were eligible for ART initiation. Baseline ($t=0$) is defined as the time of TB treatment initiation.  The total number of deaths is 541.  To avoid influence of sparsely distributed large values of ART initiation times, we administratively censor the data at 1.5 years ($t_{\text{max}}$). Of the 3302 patients in case II and 1060 patients in case IV of Figure 1, 1335 and 38 were administratively censored at $t_{\text{max}}$, respectively. 


At the time when these data were collected, baseline CD4 was a key marker used to decide ART
initiation time. To be consistent with AMPATH guidelines, we divide
baseline CD4 count into groups defined by the 
intervals $[0,50]$, $(50,200]$, $(200,350]$. 
Median weeks to ART initiation is 4 for CD4 $\leq$ 50, 8 for CD4 $\in (50,200]$ and 12  for CD4 $\in (200, 350]$.

Baseline covariates and time-varying CD4 count (1.9 CD4 measures per year per person) were used to fit the hazard models  $\lambda^A(t \cond
\ol{L}(t))$ and $\lambda^C(t \cond \ol{N}^A(t), \ol{L}(t))$
leading to
treatment and censoring weights. Each model included the main effect
of each covariate; higher-order polynomial terms for continuous variables
were tested and found not to add information. Estimated model coefficients are summarized in Web Table~2. 
Bootstrap re-sampling with 1000 replicates is used to estimate standard errors
of the estimated mean one-year survival; censoring and ART initiation time models are 
re-fit within each bootstrap sample.

All calculations are carried out in \verb+R+.  The hazard models for 
ART initiation time and censoring time are 
fitted using \verb+R+ function \verb"coxph".
The structural model given by \eqref{eq:hazardmodel3_p1}
is fitted using weighted partial likelihood, also using \verb+coxph+ \citep{survival-package},
with spline basis functions for $g_1(a)$, $g_2(t-a)$ and $g_3(a(t-a))$ generated
using the \verb+R+ function \verb+ns+. 
We place knots 
at the 25th, 50th and 75th percentile of the uncensored values of $A^*$
for $g_1 (\cdot)$; $T^*-A^*$ for $g_2 (\cdot)$; and $A^*(T^*-A^*)$ for $g_3(\cdot)$.  

\begin{figure}[H]
\centering
\includegraphics[scale=.12]{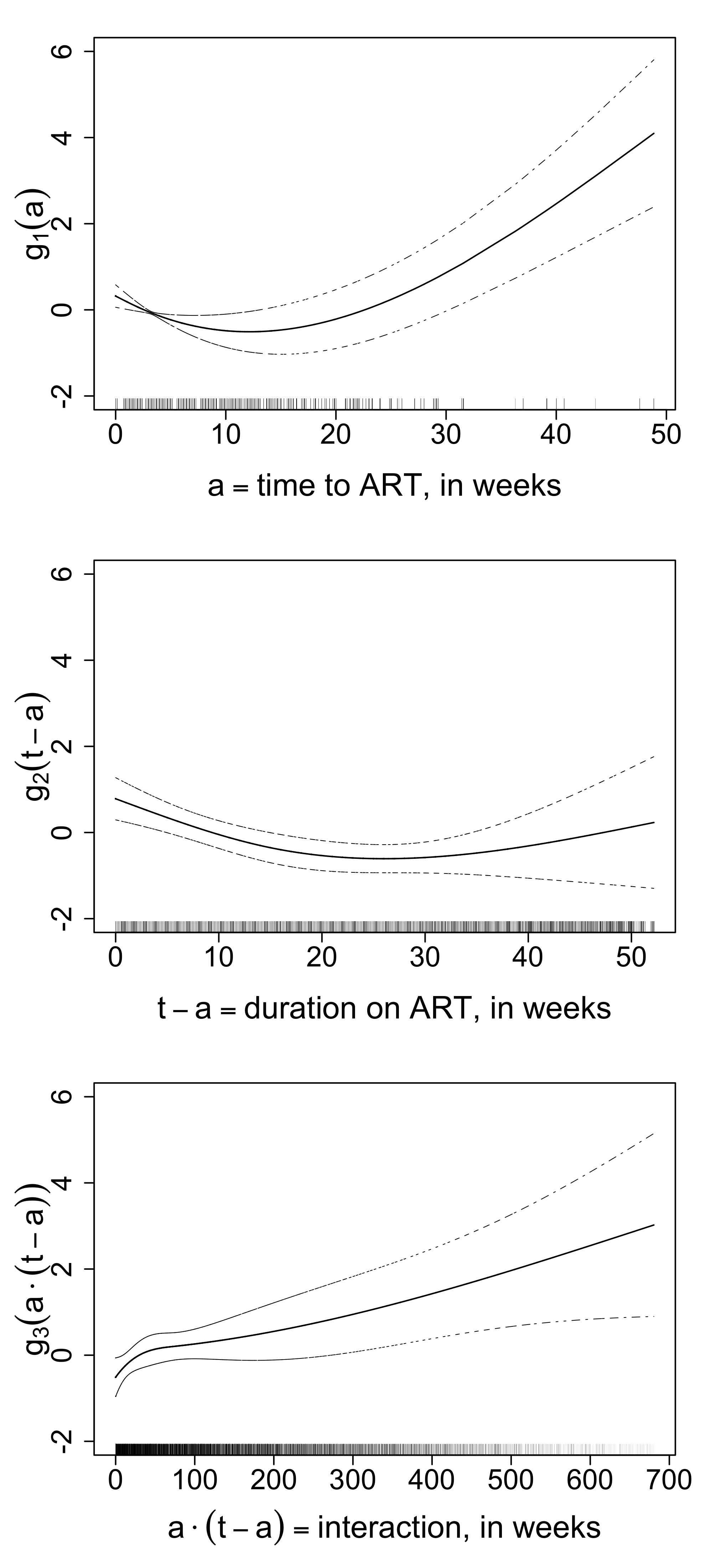}
\caption{Fitted curves $g_1(a)$, $g_2(t-a)$ and $g_3(a(t-a))$,
for baseline CD4$\leq$50. See model~\eqref{eq:rmodel}.} 
\label{fig:hplots}
\end{figure}

In Web Appendix Section 3, we describe steps of fitting the structural model~\eqref{eq:hazardmodel3_p1}, show plots of each estimated $g$ function, and summarize key findings from the plots. Plots of $g$ functions for the CD$\leq$50 subgroup appear in Figure~\ref{fig:hplots}.
\vspace{-1.2ex}
\vspace{-1.2ex}
The plot of $\wh{g}_1(a)$ in the 
first panel of Figure~\ref{fig:hplots} shows that the
instantaneous effect of ART initiation is U-shaped, with maximum benefit (lowest
mortality hazard) just after 10 weeks, and lower effectiveness with longer delays.
In the second panel, $\wh{g}_2(t-a)$ depicts the effect of treatment duration, 
and generally indicates that longer duration times
are associated with lower mortality hazard.  The increasing trend
in the interaction term
$\wh{g}_3(a(t-a))$ suggests that delayed ART initiation reduces the effect of 
duration on treatment.  The quadratic trend in $\wh{g}_1(a)$ supports the notion that immediate treatment
initiation carries some risk of elevated mortality that is possibly balanced out 
by the benefit of remaining on treatment longer.  The net causal effect is
summarized in plots comparing one-year mortality rate between regimes $a$ and $a'$,
where $a=0$ and $a'> 0$ (Figure~\ref{fig:mort.fn.a}, described below).

\begin{figure}[H]
 \centering
\includegraphics[scale=.085]{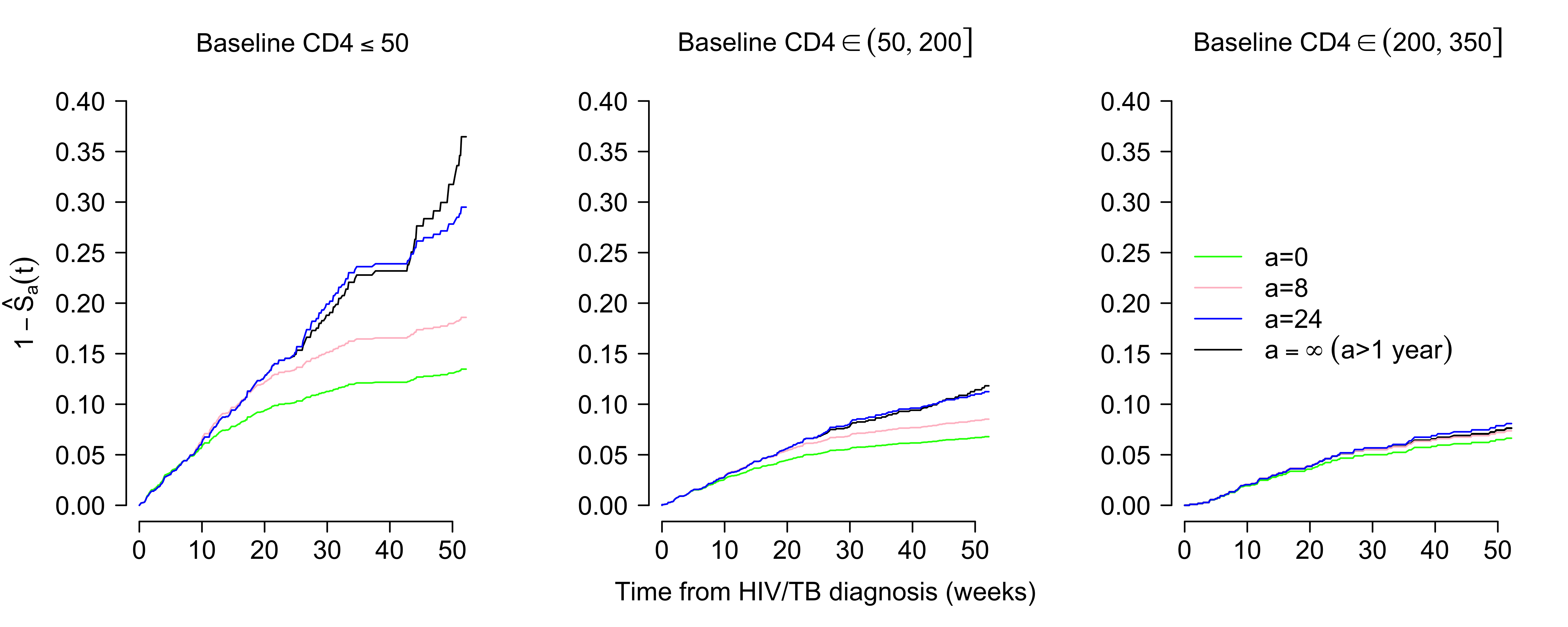}
\vspace{-.5 cm}
\caption{Estimated mortality functions $1-\widehat{S}_a(t)$
corresponding to ART initiation times $a \in \{0,\;
8,\; 24,\; \infty$\}.} 
\label{fig:surv.for.a}
\end{figure}

\begin{figure}[H]
 \centering
\includegraphics[scale=.45]{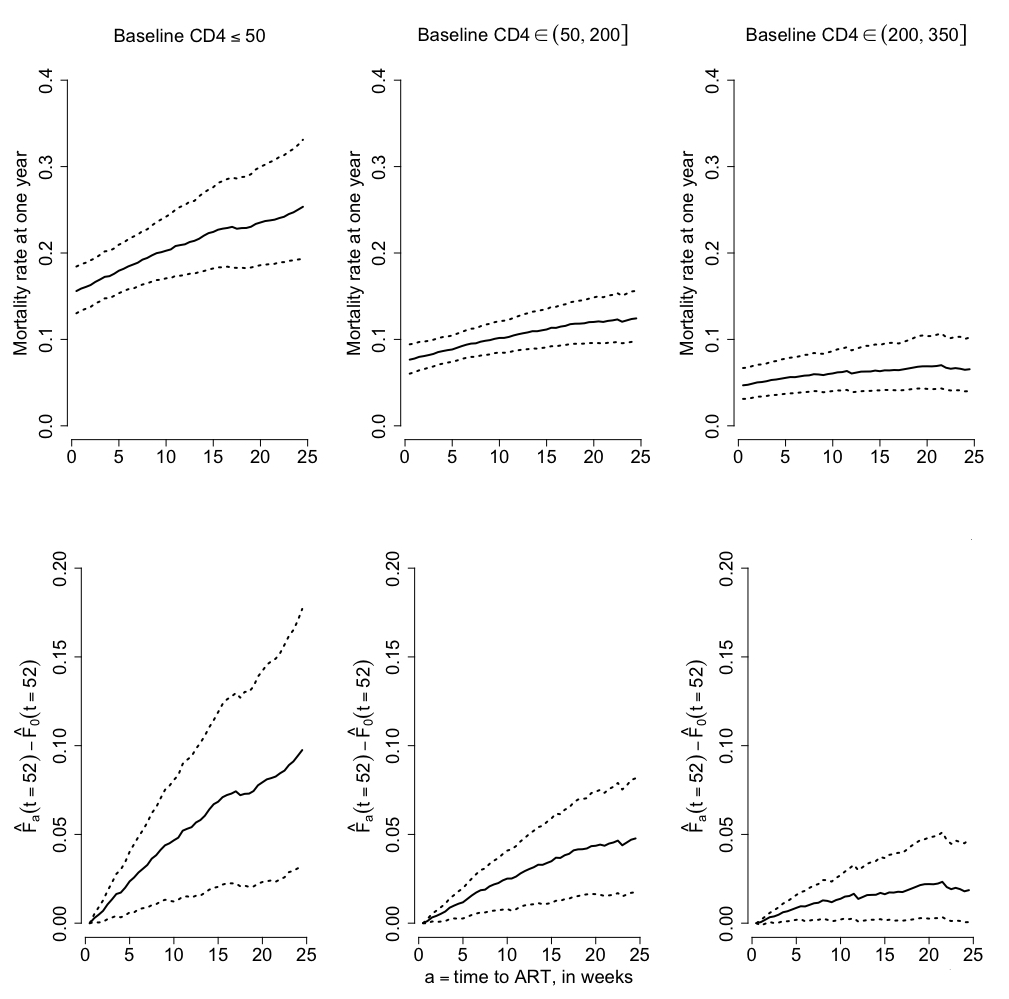}
\caption{Top panel: causal effect of ART initiation: one-year mortality as a
function of ART initiation time. Bottom panel: difference in one-year mortality between initiating ART at time $a$ and concomitant initiation with TB treatment. Stratified by baseline CD4 count.}
\label{fig:mort.fn.a}
\end{figure}

Figure~\ref{fig:surv.for.a} shows estimated mortality curves, derived using~\eqref{eq:surv}, for selected ART initiation times $a \in   \{0, 8, 24, \infty\}$ and stratified by baseline CD4 subgroup. Differences between these curves are causal contrasts. 
For each distinct value of $a$, mortality rate is clearly highest
for those with baseline CD4 $\in [0,50]$.
Looking within CD4 strata, comparing curves for $a=0$ and $a=8$ to 
those with $a > 8$ suggests a benefit of early initiation for
all groups, with greatest benefit for those in the lowest CD4 stratum. Figure~\ref{fig:mort.fn.a} shows one-year mortality and the causal treatment effect $\wh{S}_a(52) - \wh{S}_0 (52)$ as a function of treatment initiation time,  stratified by baseline CD4 subgroup. The effect of early initiation is most pronounced for  CD4 $\leq 50$. The results suggest immediate initiation for CD4$\leq 200$. Diagnostic plots suggest the model fits the data well (Web Figure~4).
\vspace{-1.2ex}
\vspace{-1.2ex}

\vspace{-1.2ex}
\vspace{-1.2ex}

Finally, we can emulate comparisons between regimens reported
in the randomized trials cited earlier. We can mimic random allocation of treatment initiation time to specific intervals by assuming a distribution for $A$ that is independent of outcomes and covariates, and compare interval-specific mortality rates  to draw inferences about treatment timing. In our data example, we assume $A$ follows a uniform distribution, and calculate  one-year survival probability associated with initiating in a given interval $[t_1, t_2)$ via $\int_{ a \in [t_1,t_2) }  \{ 1 - \wh{F}_a(t_0) \} d F^U(a)$, where $ F^U(\cdot)$ is the CDF of the uniform distribution on $ [0, t_{\text{max}}]$. To illustrate,  following the SAPIT trial, we compare one-year mortality for initiating intervals $[0,4]$ and $[8,12]$ weeks. For those with CD4 $\leq 50$ ($n=1540$),  our model estimated the mean mortality rate  to be .16 $(.14, .19)$ for $[0,4]$ and .20 $(.17, .24)$ for $[8,12]$, with a $p$-value of .02 for the difference between the means,  as compared to .10 and .20 respectively for the two regimens from the trial ($n=72$), with a $p$-value of .17 for the incidence-rate ratio. 

Our results suggest ART should be initiated within 8 weeks of the
initiation of TB therapy for AMPATH patients with CD4 counts lower
than 200. 
There is a marked increase in expected one-year mortality when ART is initiated after 8 weeks (see Figure~\ref{fig:mort.fn.a}). 
For patients with CD4 $\in (200, 350]$, while early initiation still results in lowest one-year mortality, there is not strong statistical evidence to support the conclusion that a specific
ART initiation time (or time interval) will lead to reduced one-year mortality. 
Our results are consistent with AMPATH guidelines (Web Table~3)  in treating HIV/TB co-infected patients, and are consistent with general findings of randomized controlled trials.

\section{Simulation}\label{sec:simulation}
We conducted a simulation to study properties of our weighted estimator when the model is correctly specified and to evaluate sensitivity of the weighted estimator to violations of the no unmeasured confounding assumption. We simulate data from a simplified version of our structural model~\eqref{coxmsm_p1}. 
We examine bias, variability and confidence interval coverage rates related to estimates of one-year mortality for different choices of $a$ under three scenarios: 1) under random allocation of treatment; 2) with measured confounding and 3) in the presence of unmeasured confounding.  The simulation results show near-zero bias and nominal coverage probability for Scenario 1. In the presence of measured confounding (Senario 2), the weighted estimator eliminates nearly all the bias and coverage probabilities are close to nominal levels as compared to the unweighted estimator. Unmeasured confounding (Scenario 3) produces bias in proportion to the degree of confounding. We provide details about the data generation algorithm, the parameter values and results in Web Appendix Section 2. 


\section{Summary and Discussion}

Timing of ART initiation is
important in HIV/TB co-infection. Determining the optimal ART
initiation time is difficult because of the need
to achieve balance between the risk for mortality and
potential adverse events associated with early antiviral therapy initiation.
Three randomized controlled trials (AACTG, SAPIT,
and CAMELIA) support earlier initiation of ART for those with 
very low CD4 count, but initiation times are quantified on an
interval scale.  Our data, derived from electronic health records,
allow higher-resolution inference for initiation time on a continuous
scale, are sampled from a well-defined population,
and reflect outcomes from actual clinical practice.  


Our results are largely consistent with the findings of 
the clinical trials, and provide important reinforcement and
elaboration of those findings.  
Our model provides estimates of mortality for any potential
initiation time ranging from 0
to 40 weeks, and moreover it captures the separate effects of
ART initiation and ART duration.  This latter feature enables examination
of the potential trade-offs associated with early initiation,
illustrated in Figure~\ref{fig:hplots}. 
In addition to reinforcing the findings from recent trials,
inference about grouped-interval-based optimal initiation times
for ART are consistent with current AMPATH guidelines. 

Our model assumes that both treatment initiation and
censoring are independent of potential mortality outcomes, conditionally
on baseline and time-dependent covariates.  Given that treatment
decisions are based on clinical indicators that we included in our
weight model, the treatment ignorability assumption has substantive 
justification.  It is possible that censoring could be associated with higher death rate. 
In the future, we will develop formal representations of potential selection biases, and use those as a basis for examining sensitivity to violations of assumptions about ignorable treatment assignment and censoring.

Our model, though flexible, can be extended in several directions.
First, an important structural assumption is that the effects of
initiation, duration and their interaction are additive on the log hazard scale;
this requirement could be relaxed by introducing a multi-dimensional
spline function.  Fitting this sort of model would likely require a
larger dataset. Second, our model considers regimens that are
static in the sense that they are dependent on a baseline covariate
(here, CD4 count).  A potential topic of further research is comparison 
of dynamic treatment regimes, which are defined in terms of an individual's evolving history;
e.g., `initiate ART when CD4 first drops below $x$'
\citep{robins2008estimation}.
Although we considered this approach, 
in AMPATH --- as in most resource-constrained HIV care environments ---
data needed for these comparisons are limited because
CD4 and other HIV disease markers are measured infrequently 
(typically every six months, but sometimes less often).

Finally, focused clinical questions that have generated considerable data
from both randomized trials and observational studies are potentially
fertile ground for development of new methods that combine, synthesize
or compare available evidence.  This activity has particular importance
in the field of HIV, where large observational databases are forming the
basis for important and far-reaching policy related to treatment and
policy interventions \citep{AIDSPanel2016}.  

\backmatter

\section*{Acknowledgements}
The authors are grateful to E.\ Jane Carter, MD, Rami Kantor, MD, Michael Littman, PhD, Tao Liu, PhD, Xi (Rossi) Luo, PhD (Brown University); Brent Johnson, PhD (University of Rochester),  and 
Beverly Musick, MS (Indiana University) for helpful input and discussion, and to the editor, associate editor, and two anonymous reviewers whose comments and suggestions led to substantial improvements in the manuscript. Beverly Musick also constructed
the dataset that was used for this analysis.  This work was partially funded by grants
R01-AI-108441, R01-CA-183854, U01-AI-069911, and P30-AI-42853 from the U.S.\ National Institutes of Health, and contract number 623-A-00-0-08-00003-00 from United States Agency for International Development (USAID).
\vspace*{-8pt}


\section*{Supplementary Materials}

Web Appendices, Tables, and Figures referenced in Sections~\ref{sec:estimation},~\ref{sec:application},~\ref{sec:simulation}, and the \verb+R+ code used to implement our approach are available with this paper at the Biometrics website on Wiley Online
Library. \vspace*{-8pt}


%
 \bibliographystyle{biom} 
 \bibliography{dissertation.bib}

%
%
%



%

\label{lastpage}

\end{document}